\documentclass[12pt]{article}

\usepackage[margin=1.0in]{geometry}

\usepackage{amsmath,amsfonts,amsbsy,amssymb,array,accents,dsfont}
\usepackage{enumerate}
\usepackage[shortlabels]{enumitem}
\usepackage{graphicx} 
\usepackage{caption} 
\usepackage{subcaption} 
\usepackage{mathrsfs}
\usepackage{upgreek}
\usepackage{bm}
\usepackage{slashed}
\usepackage{chngpage} 

\usepackage{color}

\usepackage{xparse}
\usepackage{xspace}

\NewDocumentCommand\eqn{om}{%
  \IfNoValueTF{#1}
     {\[ #2 \]}
     {\begin{equation}\label{#1} #2  \end{equation} \expandafter\newcommand\csname #1\endcsname{\eqref{#1}\xspace}\ignorespaces}
}
\NewDocumentCommand\eqna{om}{%
  \IfNoValueTF{#1}
    {\begin{align*} #2 \end{align*}}
    {\begin{equation}\label{#1}\begin{split} #2  \end{split}\end{equation} \expandafter\def\csname #1\endcsname{\eqref{#1}\xspace}\ignorespaces}
}

\newcommand{\rcite}{\cite}

\def\rmf{{\rm f}}
\def\rmi{{\rm i}}

\def\xx{{\bf x}}
\def\YM{{\rm \sst YM}}
\def\ptcl{{\rm ptcl}}

\def\BH{{\rm BH}}

\def\sl{\text{sl}}

\def\ptcl{{\rm brane}}

\def\BH{{\rm BH}}

\def\tight#1{\! #1 \!}  

\def\({\left(}
\def\){\right)}
\def\[{\left[}
\def\]{\right]}

\def\eg{{e.g.}}

\def\etc{{etc}}

\def\sfI{{\mathsf I}}

\def\sfm{{\mathsf m}}

\DeclareMathSymbol{\medhatsym}{\mathord}{largesymbols}{"62} 

\DeclareMathSymbol{\medtildesym}{\mathord}{largesymbols}{"65}

\makeatletter
\newcommand*\rel@kern[1]{\kern#1\dimexpr\macc@kerna}
\newcommand*\widebar[1]{%
  \begingroup
  \def\mathaccent##1##2{%
    \rel@kern{0.8}%
    \overline{\rel@kern{-0.8}\macc@nucleus\rel@kern{0.2}}%
    \rel@kern{-0.2}%
  }%
  \macc@depth\@ne
  \let\math@bgroup\@empty \let\math@egroup\macc@set@skewchar
  \mathsurround\z@ \frozen@everymath{\mathgroup\macc@group\relax}%
  \macc@set@skewchar\relax
  \let\mathaccentV\macc@nested@a
  \macc@nested@a\relax111{#1}%
  \endgroup
}
\makeatother

\def\half{\frac12}

\def\hf{\coeff12}

\def\One{{\hbox{1\kern-1mm l}}}

\def\Im{{\sfI\sfm\,}}

\def\barray{\begin{array}}
\def\earray{\end{array}}
\def\be{\begin{equation}}
\def\ee{\end{equation}}
\def\bea{\begin{eqnarray}}
\def\eea{\end{eqnarray}}
\def\bal{\begin{align}}
\def\eal{\end{align}}

\newcommand{\bS}{{\mathbb S}}
\newcommand{\bT}{{\mathbb T}}

\definecolor{cardinal}{rgb}{0.6,0,0}
\definecolor{darkgreen}{rgb}{0,0.4,0}
\definecolor{green}{rgb}{0,0.4,0}
\definecolor{golden}{rgb}{0.92, 0.7, 0}
\definecolor{midnight}{rgb}{0, 0, 0.5}
\definecolor{darkblue}{rgb}{0, 0, 0.7}

\numberwithin{equation}{section}

\mathchardef\mhyphen="2D

  \def\cF{\mathcal {F}}
\def\cG{\mathcal {G}} \def\cH{\mathcal {H}} 
  
 \def\cN{\mathcal {N}} \def\cO{\mathcal {O}}
  \def\cR{\mathcal {R}}
\def\cS{\mathcal {S}}

\def\one{{\hbox{\kern+.5mm 1\kern-.8mm l}}}
\def\zero{{\hbox{0\kern-1.5mm 0}}}

\newcommand{\ket}[1]{{\,| {#1} \rangle}}

\def\id{\textrm{id}}

\def\id{{1 \kern-.28em {\rm l}}}

\def\journal#1&#2(#3){\unskip, \sl #1\ \bf #2 \rm(19#3) }
\def\andjournal#1&#2(#3){\sl #1~\bf #2 \rm (19#3) }

\def\eg{{\it e.g.}}

\def\etc{{\it etc}}

\def\sst{\scriptscriptstyle}

\def\half{\frac12}
\def\hf{{\textstyle\half}}
\def\ket#1{|#1\rangle}

\def\One{{1\hskip -3pt {\rm l}}}

\catcode`\@=11
\def\slash#1{\mathord{\mathpalette\c@ncel{#1}}}
\def\underrel#1\over#2{\mathrel{\mathop{\kern\z@#1}\limits_{#2}}}

\catcode`\@=12

\def\ket#1{\left| #1\right\rangle}

\def\exp{{\rm exp}}

\def\eg{{\it e.g.}}

\begin{document}

\vspace{-2cm}
\begin{flushright}
{March 29, 2023}
\end{flushright}
\vspace{1cm}

\begin{adjustwidth}{-20mm}{-20mm}
\begin{center}
{\huge {\bf  
Charge currents, rare decays,\\[.3cm] 
and black holes
}}
\end{center}
\end{adjustwidth}

\vspace{1.2cm}

\begin{center}
{\large
\textsc{Emil J. Martinec}}
\vspace{8mm}

{\it 
Kadanoff Center for Theoretical Physics,\\ Enrico Fermi Institute and Department of Physics, \\
University of Chicago, 5640 S. Ellis Ave., Chicago, IL 60637-1433, USA\\
e-martinec@uchicago.edu
}
\vspace{6mm}

\end{center}

\begin{adjustwidth}{1cm}{1cm}
\vspace{0mm}
\rule{.88\textwidth}{1pt}\\
\noindent
\textsc{Abstract:}
Radiation of conserved charges carried by a black hole is a rare process, whose probability is suppressed by the change in the entropy between the initial and final states.  This universal result provides insight into the black hole's internal structure and dynamics.\\
\rule{.88\textwidth}{1pt}\\
\end{adjustwidth}
\vspace{ -.5cm}
\begin{center}
Essay written for the Gravity Research Foundation \\ 2023 Awards for Essays on Gravitation.
\end{center}
\vspace{ .5cm}




%
%


%



\vskip .5cm

\section{Introduction}
\label{sec:intro}

The realization that black holes have an entropy $S_\BH\propto 1/\hbar$, and that they radiate as black bodies at a temperature $T\propto\hbar$, has profound implications for the quantum theory of gravity.  It has proven exceedingly difficult to formulate a theory that fully explains these facts.  Attempts to quantize general relativity using standard methods of quantum field theory fails to account for the entropy, and the resulting entropy productinon seems to run afoul of at least one of the bedrock principles of locality, causality and unitarity upon which the formalism rests.

String theory has had some success explaining the entropy of near-extremal black holes, by enumerating ensembles of bound states of its various extended objects that become black holes at strong coupling.  Through the AdS/CFT correspondence, black hole thermodynamics maps to that of an equivalent dual quantum field theory; given the correspondence, for which there is ample evidence, unitarity is ensured.  But the fate of locality and causality have proven more elusive, as the inherently non-local nature of the duality map obscures our view of near-horizon dynamics and the structure of the black hole interior.

If black holes arise as brane bound states, and carry the associated brane charges, they can radiate those charges.  The emission probability is governed by the change in entropy between the initial and final states~\rcite{Keski-Vakkuri:1996wom,Parikh:1999mf,Massar:1999wg}
\be
\label{Gambh}
\Gamma \sim \exp\big[\Delta S(M,Q,J)\big] = \exp\big[S_{\rm final}-S_{\rm initial}\big]  ~.
\ee
This result is quite general~-- it applies to the radiation of angular momentum and charge (for a review, see for instance~\rcite{Vanzo:2011wq}), as well as strings~\rcite{Martinec:2023plo}; it even applies to the fragmentation of an $AdS_2$ black hole into two smaller $AdS_2$ black holes~\rcite{Maldacena:1998uz}, and so applies universally to charged and rotating black holes very near extremality, where the geometry always approaches $AdS_2$ near the horizon.  

For the creation of typical uncharged, massless Hawking quanta, one has $\Delta S = E_\ptcl/T\sim O(1)$, and a more precise calculation including the various prefactors is necessary; but for rare processes, the change in the entropy is the dominant effect.
Charge emission is such a rare event, because the presence of a macroscopic charge in the black hole enhances its entropy for a given energy above extremality, and removing charge correspondingly reduces it.  Moreover, charged Hawking quanta are massive, and as a result their emission is suppressed by an exponential factor ${\it exp}[-{m}/{T}]$; in addition, one must pay a price for the charge emission governed by the chemical potential at the horizon ${\it exp}[-{q A(r_{\rm hor})}/{T}]$.  The production rate is thus strongly suppressed, and the change in the black hole entropy is indeed the dominant effect.

In a related context, the pair production rate for charged black holes in an external field also includes a factor of the entropy for the final state black holes~\rcite{Garfinkle:1993xk,Hawking:1994ii,Brown:1994um}
\be
\label{Gamschw}
\Gamma \sim \Gamma_{\rm Schw}\, \exp\big[ S_\BH\big]
\ee
where $\Gamma_{\rm Schw}$ is the Schwinger pair production rate.  Both here and in~\eqref{Gambh}, it is natural to interpret the  factor ${\it exp}[S_{\rm final}]$ as the density of final black hole states, which are unresolved; one is computing an inclusive process.  What about the initial state entropy factor in~\eqref{Gambh}?  It plays the same role as the tunneling transition probability $\Gamma_{\rm Schw}$ does in the production of charged pairs, and so it is natural to interpret it as the typical matrix element (squared) of the emission process.


This result is compatible with the Eigenstate Thermalization Hypothesis~\rcite{Srednicki_1999}, which models matrix elements of ``simple'' operators $\cO$ in a thermal system as 
\be
\label{ETH}
\big\langle E_\rmf\big| \cO \big| E_\rmi \big\rangle = \cF_\cO(\bar E) \, \delta_{\rmf\rmi} + e^{-S(\bar E)/2} \cG_\cO(\bar E,\omega) \cR_{\rmf\rmi}
\ee
where $\bar E=\half(E_\rmf+E_\rmi)$ is the average energy, $\omega = E_\rmf-E_\rmi$ is the transition energy, and $\cR$ is a random matrix of unit variance; $\cF_\cO$ and $\cG_\cO$ are smooth functions of their respective arguments.
The result~\eqref{Gambh} indicates that this structure extends to the Hamiltonian governing the internal black hole dynamics, and its coupling to the black hole exterior.  Here we are imagining that we can approximately separate the degrees of freedom into those realizing the enormous entropy of the black hole, and the low-energy gravitational effective field theory that governs the dynamics outside the horizon.   We can then model the matrix elements for the initial black hole state $\ket{E_\rmi}$ to make a transition to the final black hole state $\ket{E_\rmf}$ plus an additional particle in the ``thermal atmosphere'' outside the black hole, via
\be
\label{BHETH}
\big\langle E_\rmf\big| \cH_{\rm int} \big| E_\rmi \big\rangle = e^{-S_\BH( E_\rmi )/2} \cG_\cH(\bar E,\omega) \cR_{\rmf\rmi}
\ee
where $\omega=E_\rmf - E_\rmi<0$ for a decay process.
Squaring and summing over final states leads to~\eqref{Gambh}.
Conversely, the adjoint matrix elements with $E_\rmf > E_\rmi$ are involved in the absorption of quanta.  Now the $e^{-S_\BH( E_\rmf )}$ suppression in the adjoint matrix element cancels against the sum over final states, and the absorption probability is order one.


Let us compute the result~\eqref{Gambh} in a well-studied class of examples, namely toroidally compactified black branes in string theory, and interpret the result.


\section{Black $D_p$-branes}
\label{sec:Dp}

Consider the example of toroidally compactified black $D_p$-branes in string theory (see~\rcite{Peet:2000hn,Lu:2009yw} for reviews, and conventions).
We can work in coordinates%
\footnote{Related to standard Schwarzschild coordinates $t,r,\Omega_{8-p}$ by the time shift 
$dt = d\tau+\sqrt{1-f/f_0}\,H^{1/2} dr/f$, and a corresponding gauge transformation of the antisymmetric tensor gauge field, such that the geometry has no coordinate singularities at the horizon.  There is nothing particularly special about these coordinates; any  coordinate system adapted to the Killing vectors that is non-singular at the horizon will do, \eg\ Eddington-Finkelstein.}
where the (string frame) supergravity fields take the form
\begin{align}
ds^2 &= H^{-1/2}\Big(- f\, d\tau^2 + d\xx_p\tight\cdot d\xx_p\Big) +2\sqrt{1-f/f_0}\, d\tau \, dr + H^{1/2} \Big(\frac{dr^2}{f_0} + r^2 d\Omega_{8-p}^2 \Big)
\\[.3cm]
f(r) &= 1-\Big(\frac{r_0}{r}\Big)^{7-p}
~~,~~~~
H(r) = 1+\Big(\frac{r_0}{r}\Big)^{7-p}\sinh^2\beta
~~,~~~~
e^{\Phi} = H^{\frac{3-p}{4}}
\\[.3cm]
C_{01...p} &= g_{s}^{-1}\coth\beta \,\big( H^{-1} -1 \big)  ~.
\end{align}
Here $f_0$ is any function regular at the horizon, for instance $f_0\tight=f\tight+{\it const.}$, and $x^i$ are periodically identified with period $R$.  Dimensionally reducing along the toroidal compactification, the black $D_p$-brane is simply a black hole charged under the gauge potential 
\be
\label{Afield}
A_0 = \int_{\bT^p} C_{01...p} \,dx^1\wedge\cdots\wedge dx^p  ~,
\ee
in a dilaton gravity theory, where the dilaton incorporates the warping of the torus volume.

The conserved charge $Q$, mass $M$ and thermodynamic quantities associated to this solution can be parametrized via
\begin{align}
\begin{split}
M &= \frac{(8-p)\Omega_{8-p} R^p }{2\mu^2 g_s^2} \,r_0^{7-p}\Big(1+\frac{7-p}{8-p} \sinh^2\beta\Big)
\quad,\qquad
T  = \frac{7-p}{4\pi r_0 \,\cosh\beta}
\\[.3cm]
Q &= \frac{(7-p)\Omega_{8-p}}{2\mu^2 g_s} \,r_0^{7-p} \sinh\beta\cosh\beta
\quad,\qquad
S = \frac{4\pi \Omega_{8-p}R^p}{2\mu^2 g_s^2}\, r_0^{8-p} \cosh\beta
\end{split}
\end{align}
where $\Omega_n=\frac{2\pi^{(n+1)/2}}{\Gamma(\half(n+1))}$ is the volume of the unit $n$-sphere, and 
$2\mu^2 = (2\pi)^7(\alpha')^4$.  Note that $Q\propto N$, the number of $D_p$-branes in the background:
\be
\Big(\frac{r_0}{\ell_s}\Big)^{7-p} \sinh\beta\,\cosh\beta =  c_p g_s N
~~,~~~~
c_p = \big(2\sqrt{\pi}\big)^{5-p} \, \Gamma\big[\hf(7-p)\big]
\ee
One has the first law of black hole thermodynamics
\be
\label{firstlaw}
dM = T \, dS + A_0(r_0) \, dQ  ~.
\ee

It is sometimes useful to work in (or near) the decoupling limit~\rcite{Maldacena:1997re,Itzhaki:1998dd} in which gravitational dynamics in the near-extremal, near-horizon regime is conjectured to be exactly dual to the $U(N)$ gauge theory which arises as the low-energy limit of the brane dynamics (for $p\le 5$).  This limit sends $\ell_s\to 0$ with the energy $RE$ and the effective gauge coupling $g_\YM R^{3-p}\sim g_s (R/\ell_s)^{3-p}$ held fixed with respect to the scale of the torus. 
For instance, for $p=3$ the limiting theory has the geometry $AdS_5\times\bS^5$ in (periodically identified) Poincare coordinates, dual to $\cN=4$ super Yang-Mills on a spatial torus.  
These decoupled theories have a rich phase structure~\rcite{Martinec:1998ja,Martinec:1999sa,Martinec:1999bf} as one dials the effective coupling and energy density above extremality.%
\footnote{In particular, the very low energy dynamics has been interpreted via dualities in terms of boosted Schwarzschild black holes in toroidally compactified M-theory~\rcite{Banks:1996vh,Banks:1997hz,Horowitz:1997fr}.  As the black brane sheds energy and charge, it will carve a path through the phase diagram.}
The proper size of the torus grows as a power of $r$ in the decoupled theory, so dimensional reduction along the torus is not particularly appropriate; one should really think of the black object as a compactified black brane rather than a black hole.

In the decoupled systems usually considered in gauge/gravity duality (for instance, in global AdS spacetimes), 
black holes do not evaporate; rather, the flux of Hawking radiation vanishes asymptotically and so the black hole sits in equilibrium in a thermal atmosphere of its own Hawking radiation.  
The usual Hawking evaporation dynamics via uncharged modes proceeds when we extract excitations from the tail of this thermal atmosphere near the spatial boundary of the decoupled geometry, either manually using boundary operators, or by restoring the asymptotically flat region, which opens the $D_p$-brane throat onto flat space at some arbitrarily large radius.  Then what were formerly bound state wavefunctions of the uncharged excitations have some small amplitude at the nexus between the top of the throat and the asymptotically flat region.  This amplitude determines the rate at which uncharged modes leak out of the throat.  But regardless of whether this standard decay channel is available, the brane bound state can disaggregate into its $N$ constituent branes in situations where a Coulomb branch of the brane dynamics is available, with the energy above extremality in the black hole converted to kinetic energy of the branes as they wander away.  By making the throat deep enough, one can ensure that charged radiation is the dominant effect, and the analysis here continues to apply even when the geometry is asymptotically flat.


\section{The Hawking process}
\label{sec:hawking}

The toroidally compactified gauge theory has a Coulomb branch along which $D_p$-branes can escape the metastable black brane bound state at finite cost in energy.  The gravitational dual of this process is Hawking radiation of quanta charged under the gauge field~\eqref{Afield}, which is often thought of as resulting from the creation of entangled brane-antibrane pairs near the horizon which gradually separate, one member of the pair escaping the black hole to spatial infinity, and the other occupying a negative energy state inside the black hole.  In the decoupled theory, uncharged Hawking quanta are bound to the vicinity of the black hole (their energy flux integrated over angular spheres vanishes asymptotically), and so charged radiation~-- the rare process that is our focus here~-- is the only available channel by which the system can decay.%
\footnote{If one restores the asymptotically flat region starting at some large radius away from the brane, then there is a competition between uncharged and charged radiation channels which one can control by varying the scale at which the crossover to flat space takes place.}

The probability amplitude for pair creation is dominated by the near-horizon region, where either the particle member of the created pair must execute a non-classical trajectory to escape from inside the horizon, or the anti-particle member of the pair must tunnel into the black hole from outside.  The tunneling amplitude can be estimated by WKB methods pioneered in~\rcite{Hartle:1976tp} and refined in~\rcite{Kraus:1994by,Kraus:1994fj,Keski-Vakkuri:1996wom,Parikh:1999mf,Massar:1999wg}.  The effective worldline action of the brane's center of mass degrees of freedom is reparametrization invariant
\be
\label{Sptcl}
\cS_{\ptcl} = \half \int \!d\xi \sqrt{\gamma}\Big[ \gamma^{-1} G_{\mu\nu}(x) \partial_\xi x^\mu \partial_\xi x^\nu -  \mu \Big]
-q \int \! A_\mu(x) dx^\mu ~;
\ee
passing to the Hamiltonian form of the action, the Hamiltonian is constrained to vanish, and the imaginary part of the reduced action $\int\! p\cdot dx$ comes from the radial contribution
\begin{align}
\label{ImS}
\Im \cS_\ptcl &= \Im \Big[ \int_{r_{\rm in}}^{r_{\rm out}} \! dr \, p_r  \Big] ~.
\end{align}
One now solves the Hamiltonian constraint for the radial momentum $p_r$ in terms of the other canonical variables.  Due to the vanishing of $g_{\tau\tau}$ at the horizon, $p_r$ has a pole at the horizon whose residue is determined entirely by the conserved charges (the energy $E$ and charge $q$ of the quantum) and their corresponding conjugate potentials, as well as the surface gravity~$\kappa$~\rcite{Keski-Vakkuri:1996wom,Martinec:2023plo}
\be
\label{prptcl}
p_r \sim \frac{E - q A_0(r_0)}{\kappa (r-r_0)}  ~.
\ee
The radial integral across the horizon is carried out by deforming the contour slightly into the complex $r$ plane to avoid the pole, and picking up half the residue.  Using the fact that the conserved charges carried away by the escaping quanta subtract from those of the remaining black hole, and applying the first law of black hole thermodynamics~\eqref{firstlaw}, one has
\be
\Im \cS_\ptcl = \half \frac{E-q A_0(r_0)}{T_\BH} = -\half \frac{\delta M_\BH-\delta Q_\BH \,A_0(r_0)}{T_\BH} = -\half \delta S_\BH  ~,
\ee
which is the advertised result~\eqref{Gambh} when we square the WKB amplitude $e^{i\cS_\ptcl }$ to evaluate the semiclassical transition probability.
This result straightforwardly generalizes to spinning black branes~\rcite{Harmark:1999xt} and the radiation of quanta carrying angular momentum, or any other conserved charge~\rcite{Martinec:2023plo}, and highlights the universality and genericity of the radiation process.

While we have not incorporated the back-reaction of the emitted quantum on the ambient near-horizon geometry, it is possible to solve the constraints of the bulk gravity theory coupled to the worldvolume dynamics of the escaping quantum~\rcite{Kraus:1994by,Kraus:1994fj,Keski-Vakkuri:1996wom,Parikh:1999mf,Massar:1999wg}.  The constraints can be boiled down to a coupling between the radial dynamics of the quantum in terms of the conjugate pair $r,p_r$ on the worldline, and the conjugate pair of gravitational variables given by the horizon area $A$ and the Rindler time $\Theta$ near the horizon.  This coupling relates the change in the horizon area to the worldline reduced action beyond the linearized level, and extends the result~\eqref{Gambh} beyond small changes in the entropy.

Once emitted via the averaged transition probability $\Gamma\sim \exp[{\delta S_\BH}]$, a Hawking quantum encounters the ambient geometry.  There is a transmission amplitude for the quantum to escape to infinity, and a reflection amplitude to be re-absorbed by the black hole.  This is the greybody filter that surrounds any black hole (see \eg~\rcite{Harmark:2007jy}), whose effects are not incorporated in the above calculation, but could be included by a more detailed analysis of the path integral for the escaping quantum as it propagates away from the very near-horizon region.  For instance, uncharged quanta in the decoupling limit lie in bound state wavefunctions, and so will always reach some maximum radius and fall back into the black hole (or if they are massless quanta, reach the spatial boundary and reflect back into the black hole).  On the other hand, charged quanta with sufficient initial radial momentum will escape onto the Coulomb branch of the $D_p$-branes and travel to spatial infinity along a timelike trajectory.


\section{Interpretation}
\label{sec:interp}

The WKB calculation of black hole radiance highlights the central role of the classical horizon; the entire emission probability (apart from greybody factors) comes from the infinitesimal neighborhood of the horizon, which is where the geometry encodes the information about the density of initial and final states.

It seems that given the validity of gauge/gravity duality, the horizon serves as a proxy in the effective theory for the unresolved black hole density of states, and is not a causal barrier per se in the theory at finite $\hbar$ where the density of states is finite.  Classical black holes are black because an incident quantum is easily absorbed into a phase space whose volume diverges in the $\hbar\to0$ limit, which it ergodically explores forever; the probability of return vanishes.  Quantum black holes are not black, and there is no causal barrier to them emitting the information stored in them; the emission probability is the essentially kinematic factor~\eqref{Gambh} embodying the likelihood that energy and charge \etc.\ stored in the black hole leaks out into the much smaller region of phase space consisting of ordinary effective field theory quanta outside the black hole.  The emission amplitude is a small tail near the horizon of the coherent wavefunction of the black hole's underlying entropic degrees of freedom.

The situation here is analogous to a flat band of microstates in a material, such as a partially filled Landau level (so that there is a large density of states).  One describes excitations of this system in terms of the creation of (quasi)particle-hole pairs.  Ignoring the state of the Fermi sea (or more to the point, lacking the energy resolution to keep track of it), one would say that the particle-hole pair is a Bell pair and that when one removes the particle from the material, it is entangled with the hole and not with the rest of the state of the material.  But the hole member of the pair is a proxy for a de-excited state of the material with one fewer electron; the emitted electron is entangled with that state.  If one were to approximate the transition matrix element with some ETH-like approximation along the lines of~\eqref{BHETH}, and trace over the unresolved final states in the flat band, one would generate a density matrix and wonder whether there was something wrong with unitarity in the dynamics of flat bands.

The black hole is no different.  The black hole with an anti-particle partner to the emitted Hawking quantum is a less excited black hole; the anti-particle is a convenient fiction for the purposes of the effective field theory calculation, and the ``horizon'' is the mechanism by which it accounts for the black hole microstates, and the location where effective field theory degrees of freedom outside the black hole gain access to those microstates.  The small emission probability~\eqref{Gambh} is essentially a consequence of the enormous phase space internal to the black hole together with the Eigenstate Thermalization Hypothesis for off-diagonal matrix elements of its dynamics.  Causal structure doesn't seem to enter into the discussion; indeed, the above results point to the black hole ``horizon'' being not so much a causal barrier as it is the portal to the black hole's huge internal phase space, traversible in both directions.   
The slow rate of emission of Hawking radiation is simply a consequence of the vast disparity between the available phase space on either side of the ``horizon''.  The ``horizon'' only becomes a causal barrier in the strict classical limit $\hbar\to 0$ where the size of the internal phase space becomes infinite.

How much ``reality'' then should we ascribe to the effective field theory description of the horizon, and the black hole interior beyond?  It is the assumption that the internal structure is well approximated by a semiclassical {\it vacuum} spacetime, in the vicinity of the horizon and for some distance beyond, that leads to the information paradox~\rcite{Hawking:1976ra}~-- particularly in its modern presentation in terms of quantum entanglement~\rcite{Mathur:2009hf}.  
The Hawking calculation (1) doesn't resolve which final state one has, (2) averages over initial states, and (3) doesn't correctly account for the change in the entropic degrees of freedom in the microstates resulting from the transition.  It erases all the entanglement between the emitted quantum and the remaining black hole, replacing it by a density matrix.   

This issue is put into stark relief by the above calculation of discharging a $D_p$-brane black hole by the emission of the underlying $D_p$-branes.  In the bulk effective field theory, the $N$ $D_p$-branes in the background travel a trajectory that is well-separated from the antibrane Hawking partners.
The Gauss law for the antisymmetric tensor gauge theory under which the branes are charged keeps track of the initial infalling branes.  The (timelike) singularity is the source of the tensor gauge flux after the initial collapse.
The Hawking process then draws brane-antibrane pairs from the vacuum near the horizon that are completely entangled with one another, and thus the emitted brane is not at all entangled or correlated with the the original branes that formed the black hole, which are macroscopically spacelike separated from this process.  The emitted brane knows nothing about the microstate of the black hole it emerged from, according to this line of reasoning.  

In the dual gauge theory, on the other hand, the emitted brane {\it is} one of the $N$ background branes in the initial bound state.  It is entangled with the remaining $N-1$ branes rather than with some antibrane vacuum fluctuation.  This suggests that   the gravitational description, rather than having vacuum structure at and just within the horizon, with the emitted branes arising from vacuum fluctuations near the horizon, instead has a wavefunction for the background branes with support at the horizon, and it is one of these branes that undergoes the tunneling process calculated above.   
If the branes emerging are the same branes that formed the black hole to begin with, then at any given moment some fraction of those branes have to be lurking near the horizon, waiting for their chance to escape their imprisonment.  And then the Gauss law detects their presence, and one does not have the effective field theory vacuum at and within the horizon.  Unitarity of charged Hawking radiation, and locality of the effective dynamics just outside the horizon, requires the charged matter emerging from the horizon to be the original branes, on-shell {\it at} the horizon, with a tiny amplitude governed by the matrix element~\eqref{BHETH}.  
The ``tunneling'' amplitude calculated above represents some average over the microstate dynamics.
The emerging brane does not come from further inside a black hole with vacuum geometry in the interior, as then the brane would traverse a longer non-classical trajectory~\eqref{ImS}, leading to a further suppression of the emission probability that is incompatible with the thermodynamics.

How can this be?  Ordinary matter is destined to collapse into a singularity, leaving vacuum geometry in its wake.  Perhaps the answer lies in the fact that the entropy in the black hole phase is associated to a deconfinement transition in the dual gauge theory~\rcite{Witten:1998zw}.  This suggests that the horizon is a phase boundary, and that rather than being in a quiescent, vacuum state, the black hole interior is in a non-geometric phase.  The gravitational degrees of freedom are a set of $O(N^0)$ collective modes of the gauge theory (associated to single-trace operators) that are swamped in the deconfined phase by the full $O(N^2)$ set of supergluons.  There are of order $1/\hbar$ more species of excitation present than supergravity accommodates.  These excitations are both the source of black brane entropy, and serve to bind the branes together, making their escape exponentially unlikely.  But the plasma of internal brane excitations is expected to be in causal contact with the collective modes describing the black brane exterior, and influence the state of the thermal atmosphere so that Hawking quanta carry away entanglement with the branes that initially formed the black hole.  The emission of a Hawking quantum de-excites the existing excited interior degrees of freedom rather than exciting new degrees of freedom that were initially in a lower excitation state, maintaining consistency with the reduced dimensionality of the state space at each step.

This picture of the dynamics of brane emission from a thermal brane bound state is the strong-coupling version of what transpires at weak coupling.  There, the branes are bound together by the gas of nonabelian gluonic excitations.  Adjoint scalars in the Yang-Mills supermultiplet in particular are all excited, and prevent any one of the branes from escaping.  On rare occasions, all the off-diagonal modes binding one of the branes to the others fluctuate into a low excitation state, allowing that brane to escape onto the Coulomb branch.  Here one certainly expects the emission rate to be governed by the sorts of generic phase space considerations of the eigenstate thermalization hypothesis, leading to matrix elements along the lines of~\eqref{BHETH} and decay rates of the sort~\eqref{Gambh}.  It is natural to suppose that the horizon scale is the scale of support of the eigenvalue distribution of the adjoint scalars in the thermally excited bound state at strong coupling, and marks the transition between the black hole's interior and exterior, with the exterior naturally associated with the Coulomb branch of the brane dynamics.

%



\vskip 1cm

\noindent{\bf Acknowledgments:}
The work of EJM is supported in part by DOE grant DE-SC0009924. 

\vskip 2cm

\bibliographystyle{JHEP}
\bibliography{fivebranes}

\providecommand{\href}[2]{#2}\begingroup\raggedright\begin{thebibliography}{10}

\bibitem{Keski-Vakkuri:1996wom}
E.~Keski-Vakkuri and P.~Kraus, \emph{{Microcanonical D-branes and back
  reaction}},
  \href{http://dx.doi.org/10.1016/S0550-3213(97)00085-0}{\emph{Nucl. Phys. B}
  {\bfseries 491} (1997) 249--262},
  [\href{https://arxiv.org/abs/hep-th/9610045}{{\ttfamily hep-th/9610045}}].

\bibitem{Parikh:1999mf}
M.~K. Parikh and F.~Wilczek, \emph{{Hawking radiation as tunneling}},
  \href{http://dx.doi.org/10.1103/PhysRevLett.85.5042}{\emph{Phys. Rev. Lett.}
  {\bfseries 85} (2000) 5042--5045},
  [\href{https://arxiv.org/abs/hep-th/9907001}{{\ttfamily hep-th/9907001}}].

\bibitem{Massar:1999wg}
S.~Massar and R.~Parentani, \emph{{How the change in horizon area drives black
  hole evaporation}},
  \href{http://dx.doi.org/10.1016/S0550-3213(00)00067-5}{\emph{Nucl. Phys. B}
  {\bfseries 575} (2000) 333--356},
  [\href{https://arxiv.org/abs/gr-qc/9903027}{{\ttfamily gr-qc/9903027}}].

\bibitem{Vanzo:2011wq}
L.~Vanzo, G.~Acquaviva and R.~Di~Criscienzo, \emph{{Tunnelling Methods and
  Hawking's radiation: achievements and prospects}},
  \href{http://dx.doi.org/10.1088/0264-9381/28/18/183001}{\emph{Class. Quant.
  Grav.} {\bfseries 28} (2011) 183001},
  [\href{https://arxiv.org/abs/1106.4153}{{\ttfamily 1106.4153}}].

\bibitem{Martinec:2023plo}
E.~J. Martinec, \emph{{The Holar Wind}},
  \href{https://arxiv.org/abs/2303.00234}{{\ttfamily 2303.00234}}.

\bibitem{Maldacena:1998uz}
J.~M. Maldacena, J.~Michelson and A.~Strominger, \emph{{Anti-de Sitter
  fragmentation}},
  \href{http://dx.doi.org/10.1088/1126-6708/1999/02/011}{\emph{JHEP} {\bfseries
  02} (1999) 011}, [\href{https://arxiv.org/abs/hep-th/9812073}{{\ttfamily
  hep-th/9812073}}].

\bibitem{Garfinkle:1993xk}
D.~Garfinkle, S.~B. Giddings and A.~Strominger, \emph{{Entropy in black hole
  pair production}},
  \href{http://dx.doi.org/10.1103/PhysRevD.49.958}{\emph{Phys. Rev. D}
  {\bfseries 49} (1994) 958--965},
  [\href{https://arxiv.org/abs/gr-qc/9306023}{{\ttfamily gr-qc/9306023}}].

\bibitem{Hawking:1994ii}
S.~W. Hawking, G.~T. Horowitz and S.~F. Ross, \emph{{Entropy, Area, and black
  hole pairs}}, \href{http://dx.doi.org/10.1103/PhysRevD.51.4302}{\emph{Phys.
  Rev. D} {\bfseries 51} (1995) 4302--4314},
  [\href{https://arxiv.org/abs/gr-qc/9409013}{{\ttfamily gr-qc/9409013}}].

\bibitem{Brown:1994um}
J.~D. Brown, \emph{{Black hole pair creation and the entropy factor}},
  \href{http://dx.doi.org/10.1103/PhysRevD.51.5725}{\emph{Phys. Rev. D}
  {\bfseries 51} (1995) 5725--5731},
  [\href{https://arxiv.org/abs/gr-qc/9412018}{{\ttfamily gr-qc/9412018}}].

\bibitem{Srednicki_1999}
M.~Srednicki, \emph{The approach to thermal equilibrium in quantized chaotic
  systems}, \href{http://dx.doi.org/10.1088/0305-4470/32/7/007}{\emph{Journal
  of Physics A: Mathematical and General} {\bfseries 32} (jan, 1999)
  1163--1175}.

\bibitem{Peet:2000hn}
A.~W. Peet, \emph{{TASI lectures on black holes in string theory}},  in
  \emph{{Theoretical Advanced Study Institute in Elementary Particle Physics
  (TASI 99): Strings, Branes, and Gravity}}, pp.~353--433, 8, 2000,
  \href{https://arxiv.org/abs/hep-th/0008241}{{\ttfamily hep-th/0008241}},
  \href{http://dx.doi.org/10.1142/9789812799630_0003}{DOI}.

\bibitem{Lu:2009yw}
J.~X. Lu and S.~Roy, \emph{{Remarks on the instability of black Dp-branes}},
  \href{http://dx.doi.org/10.1016/j.physletb.2010.02.066}{\emph{Phys. Lett. B}
  {\bfseries 686} (2010) 254--258},
  [\href{https://arxiv.org/abs/0911.3341}{{\ttfamily 0911.3341}}].

\bibitem{Maldacena:1997re}
J.~M. Maldacena, \emph{{The large N limit of superconformal field theories and
  supergravity}}, {\emph{Adv. Theor. Math. Phys.} {\bfseries 2} (1998)
  231--252}, [\href{https://arxiv.org/abs/hep-th/9711200}{{\ttfamily
  hep-th/9711200}}].

\bibitem{Itzhaki:1998dd}
N.~Itzhaki, J.~M. Maldacena, J.~Sonnenschein and S.~Yankielowicz,
  \emph{{Supergravity and the large N limit of theories with sixteen
  supercharges}},
  \href{http://dx.doi.org/10.1103/PhysRevD.58.046004}{\emph{Phys. Rev. D}
  {\bfseries 58} (1998) 046004},
  [\href{https://arxiv.org/abs/hep-th/9802042}{{\ttfamily hep-th/9802042}}].

\bibitem{Martinec:1998ja}
E.~J. Martinec and V.~Sahakian, \emph{{Black holes and the superYang-Mills
  phase diagram. 2.}},
  \href{http://dx.doi.org/10.1103/PhysRevD.59.124005}{\emph{Phys. Rev. D}
  {\bfseries 59} (1999) 124005},
  [\href{https://arxiv.org/abs/hep-th/9810224}{{\ttfamily hep-th/9810224}}].

\bibitem{Martinec:1999sa}
E.~J. Martinec and V.~Sahakian, \emph{{Black holes and five-brane
  thermodynamics}},
  \href{http://dx.doi.org/10.1103/PhysRevD.60.064002}{\emph{Phys. Rev. D}
  {\bfseries 60} (1999) 064002},
  [\href{https://arxiv.org/abs/hep-th/9901135}{{\ttfamily hep-th/9901135}}].

\bibitem{Martinec:1999bf}
E.~J. Martinec, \emph{{Black holes and the phases of brane thermodynamics}},
  in \emph{{NATO Advanced Study Institute: TMR Summer School on Progress in
  String Theory and M-Theory}}, pp.~117--145, 5, 1999,
  \href{https://arxiv.org/abs/hep-th/9909049}{{\ttfamily hep-th/9909049}}.

\bibitem{Banks:1996vh}
T.~Banks, W.~Fischler, S.~H. Shenker and L.~Susskind, \emph{{M theory as a
  matrix model: A Conjecture}},
  \href{http://dx.doi.org/10.1103/PhysRevD.55.5112}{\emph{Phys. Rev. D}
  {\bfseries 55} (1997) 5112--5128},
  [\href{https://arxiv.org/abs/hep-th/9610043}{{\ttfamily hep-th/9610043}}].

\bibitem{Banks:1997hz}
T.~Banks, W.~Fischler, I.~R. Klebanov and L.~Susskind, \emph{{Schwarzschild
  black holes from matrix theory}},
  \href{http://dx.doi.org/10.1103/PhysRevLett.80.226}{\emph{Phys. Rev. Lett.}
  {\bfseries 80} (1998) 226--229},
  [\href{https://arxiv.org/abs/hep-th/9709091}{{\ttfamily hep-th/9709091}}].

\bibitem{Horowitz:1997fr}
G.~T. Horowitz and E.~J. Martinec, \emph{{Comments on black holes in matrix
  theory}}, \href{http://dx.doi.org/10.1103/PhysRevD.57.4935}{\emph{Phys. Rev.
  D} {\bfseries 57} (1998) 4935--4941},
  [\href{https://arxiv.org/abs/hep-th/9710217}{{\ttfamily hep-th/9710217}}].

\bibitem{Hartle:1976tp}
J.~B. Hartle and S.~W. Hawking, \emph{{Path Integral Derivation of Black Hole
  Radiance}}, \href{http://dx.doi.org/10.1103/PhysRevD.13.2188}{\emph{Phys.
  Rev. D} {\bfseries 13} (1976) 2188--2203}.

\bibitem{Kraus:1994by}
P.~Kraus and F.~Wilczek, \emph{{Selfinteraction correction to black hole
  radiance}}, \href{http://dx.doi.org/10.1016/0550-3213(94)00411-7}{\emph{Nucl.
  Phys. B} {\bfseries 433} (1995) 403--420},
  [\href{https://arxiv.org/abs/gr-qc/9408003}{{\ttfamily gr-qc/9408003}}].

\bibitem{Kraus:1994fj}
P.~Kraus and F.~Wilczek, \emph{{Effect of selfinteraction on charged black hole
  radiance}}, \href{http://dx.doi.org/10.1016/0550-3213(94)00588-6}{\emph{Nucl.
  Phys. B} {\bfseries 437} (1995) 231--242},
  [\href{https://arxiv.org/abs/hep-th/9411219}{{\ttfamily hep-th/9411219}}].

\bibitem{Harmark:1999xt}
T.~Harmark and N.~A. Obers, \emph{{Thermodynamics of spinning branes and their
  dual field theories}},
  \href{http://dx.doi.org/10.1088/1126-6708/2000/01/008}{\emph{JHEP} {\bfseries
  01} (2000) 008}, [\href{https://arxiv.org/abs/hep-th/9910036}{{\ttfamily
  hep-th/9910036}}].

\bibitem{Harmark:2007jy}
T.~Harmark, J.~Natario and R.~Schiappa, \emph{{Greybody Factors for
  d-Dimensional Black Holes}},
  \href{http://dx.doi.org/10.4310/ATMP.2010.v14.n3.a1}{\emph{Adv. Theor. Math.
  Phys.} {\bfseries 14} (2010) 727--794},
  [\href{https://arxiv.org/abs/0708.0017}{{\ttfamily 0708.0017}}].

\bibitem{Hawking:1976ra}
S.~W. Hawking, \emph{{Breakdown of Predictability in Gravitational Collapse}},
  \href{http://dx.doi.org/10.1103/PhysRevD.14.2460}{\emph{Phys. Rev. D}
  {\bfseries 14} (1976) 2460--2473}.

\bibitem{Mathur:2009hf}
S.~D. Mathur, \emph{{The Information paradox: A Pedagogical introduction}},
  \href{http://dx.doi.org/10.1088/0264-9381/26/22/224001}{\emph{Class. Quant.
  Grav.} {\bfseries 26} (2009) 224001},
  [\href{https://arxiv.org/abs/0909.1038}{{\ttfamily 0909.1038}}].

\bibitem{Witten:1998zw}
E.~Witten, \emph{{Anti-de Sitter space, thermal phase transition, and
  confinement in gauge theories}},
  \href{http://dx.doi.org/10.4310/ATMP.1998.v2.n3.a3}{\emph{Adv. Theor. Math.
  Phys.} {\bfseries 2} (1998) 505--532},
  [\href{https://arxiv.org/abs/hep-th/9803131}{{\ttfamily hep-th/9803131}}].

\end{thebibliography}\endgroup

\end{document}